\documentclass[aps,twocolumn,superscriptaddress, floatfix]{revtex4}
\usepackage{amsmath}
\usepackage{graphicx}
\usepackage{epsfig}

\begin{document}
% You should use BibTeX and revtex.bst for references
\bibliographystyle{apsrev}
%\preprint{}

\title{Spin and Polarized Current from Coulomb Blockaded Quantum Dots }

\author{R. M. Potok}
\affiliation{Department of Physics, Harvard University, Cambridge,
Massachusetts 02138}
\author{J. A. Folk}
\affiliation{Department of Physics, Harvard University, Cambridge,
Massachusetts 02138} \affiliation{Department of Physics, Stanford
University, Stanford, California 94305}
\author{C. M. Marcus}
\affiliation{Department of Physics, Harvard University, Cambridge,
Massachusetts
02138}
\author{V. Umansky}
\affiliation{Braun Center for Submicron Research, Weizmann
Institute of Science, Rehovot, 76100, Israel}
\author{M. Hanson}
\affiliation{Materials Department, University of California, Santa
Barbara, Santa Barbara, California 93106}
\author{A.C. Gossard}
\affiliation{Materials Department, University of California, Santa
Barbara, Santa Barbara, California 93106}
%\date{\today}

\begin{abstract}
We report measurements of spin transitions for $GaAs$ quantum dots
in the Coulomb blockade regime, and compare ground and excited state transport spectroscopy to direct measurements
of the spin polarization of emitted current. Transport spectroscopy
reveals both spin-increasing and spin-decreasing transitions as
well as higher-spin ground states, and allows g-factors to be measured down to a single electron. The spin of
emitted current in the Coulomb blockade regime, measured using spin-sensitive electron focusing, is found to be
polarized along the direction of the applied magnetic field
regardless of the ground state spin transition.
\end{abstract}
\maketitle

Quantum dots in the Coulomb blockade (CB) regime have for several
years provided a valuable tool to study spin in confined systems.
Systems with small interactions, such as nanotubes \cite{Cobden}
and nonmagnetic metal grains \cite{Ralph95}, show signatures of 
spin degenerate
orbital levels with electrons filling in a simple Pauli scheme of
spin $0, \frac{1}{2}, 0, \frac{1}{2}$, \ldots In contrast,
recent transport measurements in
lateral $GaAs$ quantum dots \cite{DGG, Folk00, Ensslin} suggest the
existence of higher-spin ground states.

\begin{figure}[!]
\includegraphics[width=3.25in]{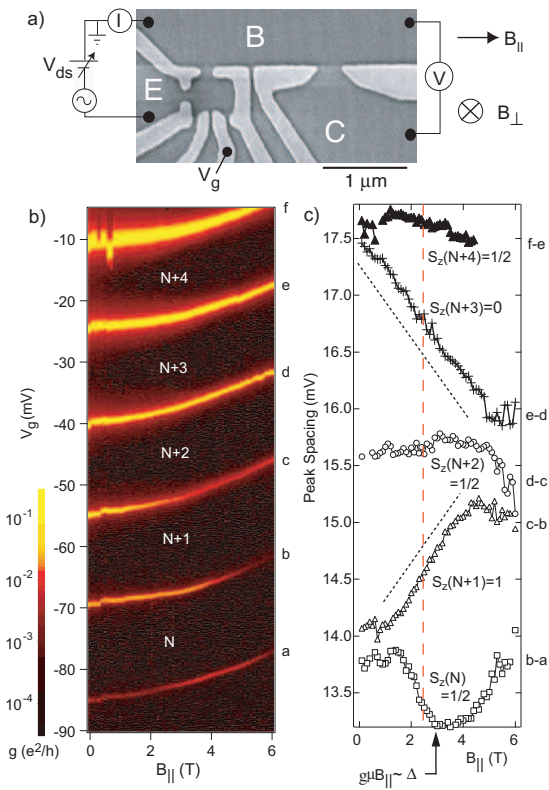}
    \caption{\footnotesize {(a) Micrograph of a quantum dot, similar to the one measured,
    in a focusing geometry.  A voltage is applied from emitter (E) to
    base (B) regions; emitter current and base-collector (B-C) voltage give dot conductance and focusing
    signal respectively.  (b) Six consecutive Coulomb blockade
    peaks in the weak tunneling regime (valley conductance near zero),
    measured as a function of gate voltage, $V_g$, and in-plane magnetic field,
    $B_{\parallel}$.  A hall bar fabricated on the same chip allows the
    perpendicular field, $B_{\perp}$, to be measured simultaneously and
    held at $\sim -110 mT$. (c) Peak spacings (in $V_g$) extracted from
    the data in (b).  From the
    slopes of these lines in $B_{\parallel}$, the spin transition associated
    with each Coulomb blockade peak may be determined.  For example,
    at $ B_{\parallel}=2.5T$ (red dashed line) a possible sequence of ground spin
    states resulting from these transitions is shown.  The
    dotted black lines indicate expected slopes of peak spacing for
    $S_{z}(N)\rightarrow S_{z}(N) \pm \frac{1}{2}$ transitions, using
    $g=0.44$.  Spacings offset for clarity.}}
\end{figure}

In this Letter, we explore ground and excited spin states of few- and
many-electron lateral $GaAs$ dots in the weak tunneling regime, using both
transport spectroscopy as well as a focusing measurement that allows a
direct determination of the spin polarization of emitted current \cite{Potok02}.  Consistent with previous work
\cite{DGG, Folk00, Ensslin} we find, as evidence of higher-spin ground states in the larger dot, that spin transitions
(increasing or decreasing) are often followed by a second transition in the same direction as electrons are added to
the dot.  Excited state spin transitions and spin degeneracy for
several quantum levels are also explored using nonlinear bias
spectroscopy,  and clear spin splitting is found for the N=1 electron case in the few-electron
dot.  It is generally believed \cite{Loss} that opposite state spin transitions lead to opposite spin
polarizations of the emitted current on Coulomb blockade peaks.  We find instead that the spin polarization of the
current is the same for CB peaks corresponding to spin-increasing and spin-decreasing transitions, with the
polarization always aligned with the external magnetic field.

Measurements were performed on two quantum
dots, one with many electrons ($N\sim100$) and the
other with few electrons ($N<10$).  In the small dot we concentrate on the $N=0 \rightarrow 1$ electron
transition.  Focusing measurements of spin polarization of emitted current were
performed for the larger quantum dot. The devices were fabricated using $Cr/Au$ depletion
gates on the surface of a $GaAs/Al_{x}Ga_{1-x}As$ heterostructure; the two dimensional electron gas (2DEG) at the interface was contacted electrically using nonmagnetic $PtAuGe$
ohmics.  For the larger dot (Fig.~1(a)) we used a heterostructure $(x = 0.36)$
with the 2DEG lying 102 $nm$ from the
surface and with electron density $n=1.3\times
10^{11} cm^{-2}$.  The high mobility of this 2DEG, $\mu = 5.5\times 10^6
cm^2/Vs$, allowed the observation of several clear focusing peaks.
Characteristic energy scales for the larger quantum dot include a level spacing $\Delta \sim 70 \mu eV$ and a
charging energy $E_c\sim 800 \mu eV$.  The smaller quantum dot (Fig.~2(b), inset \cite{Ciorga}) was
fabricated on a different heterostructure $(x = 0.3)$ with density
$2.3\times 10^{11} cm^{-2}$; the mobility was $5\times 10^{5} cm^2/Vs$.

Experiments were carried out in a dilution refrigerator with
base electron temperature $T_e=70 mK$ (determined by CB peak
width), using standard ac lock-in techniques with an excitation
voltage of $5 \mu V$. A pair of tranverse superconducting magnets was used to
provide independent control of
field in the plane of ($B_{\parallel}$) and perpendicular to
($B_\perp$) the 2DEG \cite{Folk01}.

On a CB peak, transport through an $N$-electron dot occurs via the addition
and removal of the $N+1$ electron, with the
corresponding z-component of the dot spin,
$S_{z}(N)$, changing to $S_{z}(N+1)$ and back again. The energy required for
this transition as measured by CB peak position depends on
the the magnetic field $B$ through a Zeeman term, $-g\mu B (S_{z}(N+1)-
S_{z}(N)) = -g\mu B  (\Delta S_z)$. The spacing between $N \rightarrow N+1$ and
$N+1 \rightarrow N+2$ CB peaks is given by $-g\mu B [(S_{z}(N+2) -
S_{z}(N+1))-(S_{z}(N+1) - S_{z}(N))]$.  (The effect of the magnetic field on the orbital energies is
minimized in this experiment by changing only the in-plane componenent, $B_{\parallel}$.) A CB peak position that
moves upward in gate voltage (upward in the energy required to add an additional electron) as a function of field
indicates a spin-decreasing transition; downward motion in gate voltage indicates a spin-increasing transition. In terms
of peak spacings, a spin-increasing transition of
$\Delta S_z$ followed by an spin-decreasing transition of $-\Delta S_z$ yields a spacing that increases with
field; for the opposite sequence, the peak spacing decreases with field. For the case of 
$\Delta S_z=
\frac{1}{2}$ transitions, the slopes of the spacings will be $\pm g\mu$. Consecutive transitions of the same
magnitude and in the same direction, for instance $S_z = 0 \rightarrow \frac{1}{2}\rightarrow 1$, yield a peak
spacing that does not change with field.

Six consecutive CB peaks as a function of magnetic field for the larger dot are shown
in Fig.~1(b). The parabolic dependence of peak position on $B_{\parallel}$ is believed to result from
the effect of the field on the well confinement
potential \cite{DGG, Weis}; this effect gives the same shift for all CB peaks, and
so disappears when the peak spacing is extracted. Corresponding CB spacings, shown in Fig.~1(c), display
linear motion with slopes $\pm g\mu$ and zero, where the $g$-factor is consistent with the bulk value
for $GaAs$, $g=-0.44$. 

Beginning from an arbitrary value of spin for the $N$ electron dot, $S_{z}(N)$,
we can enumerate the ground state spin transitions for the dot as additional
electrons are added (peak spacings provide no information on the
absolute magnitude of spin, only spin transitions). For example,
in Fig.~1(c) at $2.5T$, the spacing for the two peaks at the most negative gate voltage (fewest
electrons) decreases with
$B_{\parallel}$, suggesting that $S_{z}(N+1) = S_{z}(N) + \frac{1}{2}$ and $S_{z}(N) = S_{z}(N-1) - \frac{1}{2}$.  Taking
$S_{z}(N) =
\frac{1}{2}$ gives a spin structure for the states shown in Fig.~1 (labelled $N-1, N, ..., N+5$) of
($1,
\frac{1}{2}, 1,\frac{1}{2}, 0, \frac{1}{2}, 1)$ at
$B=2.5T$.  The occurrence of peak spacings with zero slope is evidence of higher-spin ground
states.  We note that no two consecutive
spacings both have slopes $+g\mu$ or $-g\mu$. This
indicates that spin changes of $\frac{3}{2}$ or greater upon adding an electron are not seen.  (Due to the negative
g-factor in GaAs, the lower-energy spin state for a single electron will generally be anti-aligned with an external
magnetic field; therefore we will define $S_z=+\frac{1}{2}$ to be anti-aligned with the field, and for consistency the
reader may then use a positive g-factor for energy calculations.)

Excited state spin transitions can be observed using finite dc drain-source bias,
$V_{ds} > g \mu B$. A change in spin between two states (either ground or excited) of the $N$ and
$N+1$ electron systems would be expected to cause the corresponding peak in differential conductance to
shift with $B$
\cite{Cobden, Ralph95}. Furthermore, any transition which is spin degenerate at $B = 0$
should split as a function of field. Excited state transitions from
several consecutive Coulomb blockade peaks in the larger dot are shown at
$V_{ds}=400 \mu V$ as a function of $B$ and $V_g$ in Fig.~2(a).
Splitting of excited state features with field is only occasionally observed, suggesting a lack of spin
degeneracy for many of these transitions.  At the same time, some distinct transitions move toward or away
from each other with slopes $\pm g\mu$, possibly indicating differences in dot spin for initial and
final states.

To eliminate the complicating effects of a many-electron system, we also measured spin transitions for the
$N=0\rightarrow 1$ electron transition using the smaller dot (Fig.\ 2(b), inset). Finite
drain-source measurements were used to find the
$0\rightarrow 1$ electron transition, see Fig.~2(b) \cite{Kouwenhoven}.  This transition displays clear
splittings for both the ground and first excited states
(Fig.~2(c)), with $g$-factors measured to be $g\sim0.37$. When more electrons were added to the device
(for example, for the $1 \rightarrow 2$ electron transition or even
more clearly for $2 \rightarrow 3$ or higher transitions)
splittings were only occasionally observed (data not shown). The simpler behavior for the $0
\rightarrow 1$ electron transition may indicate the
important effect of interactions on the spin structure of multi-electron dots \cite{das Sarma}.

\begin{figure}[t!]
          \label{fig2}
          \includegraphics[width=3.25in]{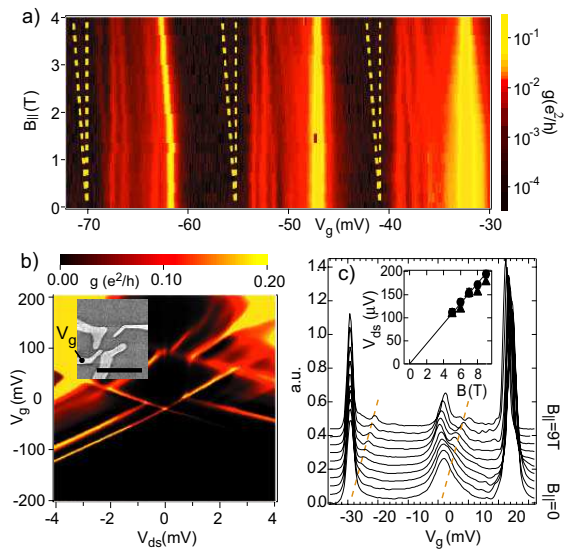}
\caption{\footnotesize {(a) Color plot of the differential
conductance of Coulomb blockade peaks at $V_{ds}=400 \mu V$, as a
function of $V_g$ and $B_{\parallel}$ ($B_{\perp}$ held constant
at $-110mT$) for the quantum dot shown in Fig.~1. (All $V_g$ traces 
were shifted to align the rightmost peak.)  For comparison
the dashed lines show an energy separation of $g\mu B$, taking $g=0.44$. Splitting is only
occasionally observed. (b) and (c) Similar measurements taken on a different quantum dot (micrograph
shown in Fig.~2(b) inset, scale bar is $1\mu m$) (b) Coulomb diamond at
$B_{\parallel}=0$ and $B_{\perp} = -200mT$ demonstrating that the
CB peak near $V_g = 0$ is the $0 \rightarrow 1$ electron
transition. (c) Differential conductance of the $0 \rightarrow 1$
electron CB peak at $V_{ds}=1200 \mu V$ from $B_{\parallel}=0$ to $9T$
(curves offset for clarity, and individually rescaled to have a constant height for the
rightmost peak). In contrast to (a), clear spin splitting of ground and excited states is seen
for this transition (dashed yellow lines are guides to the eye). Inset: splitting as a function of
$B$ for the ground state (solid circles) and first
excited state (solid triangles). Solid line shows best fit to the
data, and gives a $g$-factor of 0.37.}}

\end{figure}

In the absence of spin blockade \cite{Weis, Weinmann}, one would expect $S_z$ of the dot to change by
the the spin 
$s_z =
\pm
\frac{1}{2}$ of the electron added to it: $S_{z}(N+1) = S_{z}(N)+s_z$.
This would imply opposite polarization of transport current for
spin-increasing and spin-decreasing transitions \cite{Loss}. We examine this expectation
experimentally by comparing the spin transitions determined by CB peak position to a direct measurement
of the spin polarization of current emitted on a CB peak. 

The spin polarization of current from the quantum dot
was measured in a transverse focusing geometry (Fig.\ 1(a)).  As
described previously \cite{Potok02, Folk03}, the height of a
focusing peak reflects the degree (and direction) of spin
polarization of current from the emitter when the collector
QPC is spin selective, according to the relation
$V_c = \alpha I_e (h/2e^2)(1 + P_eP_c)$.  
Here $V_c$ is the focusing peak height, $I_e $ is the total
emitter current with polarization $P_e = (I_{\uparrow e}
- I_{\downarrow e})/(I_{\uparrow e}+I_{\downarrow e}$), and $P_c =
(T_{\uparrow c}-T_{\downarrow c})/(T_{\uparrow c}+T_{\downarrow
c})$ is the spin selectivity of the collector. (The efficiency
parameter $\alpha$ $(0 < \alpha < 1)$ accounts for
spin-independent imperfections in the focusing process.)

Using a
Coulomb blockaded quantum dot as the emitter favors the use of a voltage bias between emitter and
base, rather than a current bias as used in Refs.\ \cite{Potok02, Folk03}. In this case, changes in the
emitter current, $I_e$, lead to changes in the focusing peak height even when its polarization remains
constant.  To study spin polarization, we measure the emitter current along with the collector
voltage (Figs.~3(a) and 3(b)) and use the quantity
$V_c/I_e$, a {\it nonlocal} resistance, as a measure of the spin
polarization of the current from the CB quantum dot when the
collector is spin selective. For a spin-selective collector ($g_c= 0.5e^2/h$, in an in-plane field),
the value of 
$V_c/I_e$ should range from twice the value found in the unpolarized case ($g_c= 2e^2/h$),
when emitter polarization and collector selectivity are oriented in the same direction, to zero,
when the spin directions are oppositely oriented.

\begin{figure}[!]
          \label{fig3}
      \includegraphics[width=3.25in]{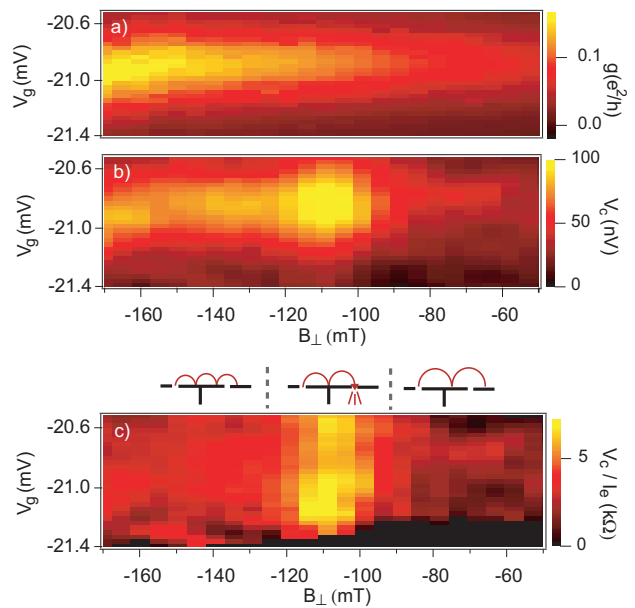}
\caption{\footnotesize {(a) Conductance of a CB peak as a function
of both $V_g$ and $B_{\perp}$, for the dot shown in Fig.~1(a) in a focusing geometry.  (b)
Base-Collector voltage, $V_c$, measured at the same time as the dot conductance, with $B_{\perp}=-110mT$ set
to correspond to the second focusing peak (the second peak was used because it was affected least by
$B_{\parallel}$ in this device). (c)  The nonlocal resistance
$V_c/I_e$ most clearly shows the effect of focusing.  The
diagrams indicate the electron focusing condition for fields near
the second focusing peak. The location of the focusing peak in
$B_{\perp}$ remained constant for all CB peaks studied.  Data does not appear when $g_e<0.1e^2/h$
($I_e<20pA$, $V_c\lesssim40nV$) because the ratio
$V_c/I_e$ becomes unreliable.}}
\end{figure}

Simultaneous focusing and conductance measurements at ${B_\parallel} = 6T$ for both spin-selective and
spin-independent collector are presented in Figs.~4(a,b), as the dot is tuned from the semi-open
to the weak tunneling regimes using the voltage, $V_g$, on the side gate.  We find that the
focusing signal $V_c/I_e$ with spin-selective collector ($g_c=0.5e^2/h$) always lies above the signal
with spin-independent collector $(g_c = 2e^2/h)$ once the dot is tuned into
the weak tunneling regime.  This suggests that
the current emitted from the quantum dot at low conductance is always spin polarized in the same
direction as the collector, over a range of gate voltage where many electrons are added.

\begin{figure}[h]
          \label{fig4}
        \includegraphics[width=3.25in]{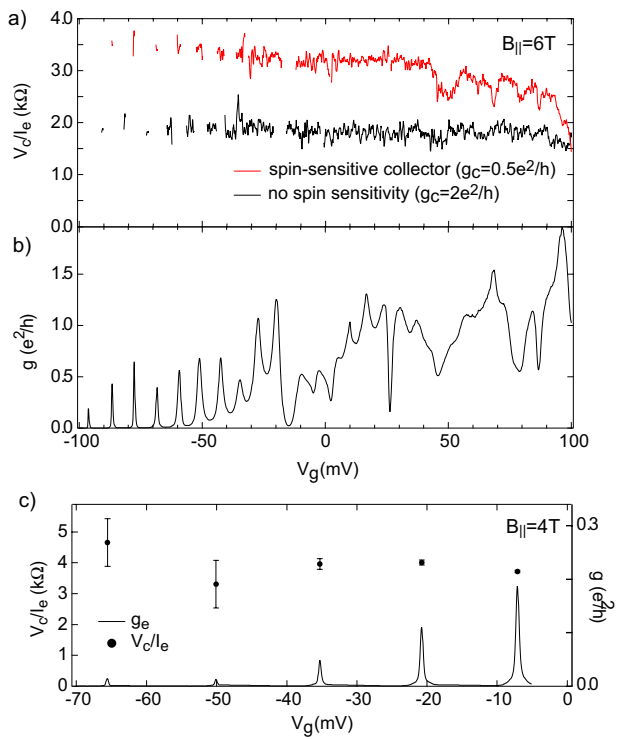}
\caption{\footnotesize {(a) Focusing signal at $B_{\parallel} =
6T$ from the quantum dot shown in Fig.~1, with spin-selective ($g_c
= 0.5e^2/h$, red curve) and spin-independent ($g_c = 2e^2/h$, black
curve) collector. The polarization of current fluctuates on a
typical gate voltage scale of $V_g=5mV$, but these fluctuations are suppressed as
$V_g$ is reduced below $30mV$. At the same
time, the spin selective curve rises to nearly twice the value as
the curve at $g_c = 2e^2/h$, indicating spin polarization of emitter current (see text).  (b) Conductance measured
simultaneously with data in (a). (c) Focusing signal and conductance measured for the CB peaks shown in Fig.~1 $(N+1$ to
$N+6)$ at
$B_{\parallel} = 4T$ and $g_c = 0.5 e^2/h$. Again, only small fluctuations in focusing signal are
observed despite different spin transitions observed for these peaks in Fig.~1. Based on the increase
of $V_c/I_e$ to $3.5k\Omega$ from
$1.9k\Omega$ with the spin selective collector in (a), we would have expected the
focusing peak to be suppressed to $ V_c/I_e \sim 0.3k\Omega$ if the opposite polarization were generated at the
emitter. (Collector selectivity depends only weakly on $B$ at these fields and temperatures
\cite{Potok02}.)}}
\end{figure}

Figure 4(c) shows focusing measurements for the same peaks shown Fig.~1,  at
$B_{\parallel} = 4T$.  Spin transitions of both directions were observed based on peak motion (see
Fig.~1) whereas spin polarization of emitted current is again found to remain nearly constant over
all measured CB peaks.  This observation is inconsistent with the picture of spin transitions leading
to
$S_{z}(N+1) = S_{z}(N)+s_z$ discussed earlier.  

We note as well that there is no apparent correlation between peak height and spin transition in a large
in-plane field.  It was shown in Refs.\ 
\cite{Potok02} and \cite{Folk03} that the leads of a quantum dot become spin polarized in the same way
as single QPC's in an in-plane field. However, a spin dependent tunnel barrier should lead to a
dramatic suppression in CB peak height for spin-decreasing transitions.  As seen in Fig.~1, this was
not observed in our measurement.  Taken together, these observations may indicate that spin
polarization in the leads is playing a role in the spin state of the quantum dot on a CB peak.

In conclusion, we have found signatures of
spin-increasing and spin-decreasing transitions in transport measurements, including spin splitting of the $N=0
\rightarrow 1$ transition. Measurements of polarization of the current emitted from a quantum dot in the CB
regime show that the emitted current is in all cases polarized in the same direction as the QPC
collector, for both spin-increasing and spin-decreasing transitions of the dot.  These observations
necessitate a revised picture of spin transitions in lateral quantum dot in an in-plane magnetic field.

We acknowledge valuable discussions with P.~Brouwer. This
work was supported in part by The Darpa SpinS and QuIST programs
and the ARO-MURI DAAD-19-99-1-0215. JAF acknowledges partial
support from the Stanford Graduate Fellowship; RMP acknowledges
support as an ARO Graduate Research Fellow.

% Create the reference section using BibTeX

\footnotesize{
%\bibliography{josh}

}

\end{document}